%% file: ECE5520.tex
\documentclass[conference]{IEEEtran}
\ifCLASSINFOpdf
   \usepackage[pdftex]{graphicx}
   \graphicspath{{../pdf/}{../jpeg/}}
\else
\fi
\usepackage{hyperref}
\usepackage{ltablex}
\usepackage{multirow}
\usepackage{tabularx}
\usepackage[color = green!50]{todonotes}
\begin{document}
%
\title{Dimming Down LED: An Open-source Threshold Implementation on Light Encryption Device (LED) Block Cipher}

\author{\IEEEauthorblockN{Yuan Yao$^*$, Mo Yang$^*$, Pantea Kiaei$^\dagger$, Patrick Schaumont$^\dagger$ }
\IEEEauthorblockA{ \\$^*$Virginia Tech, Blacksburg, VA, USA \\ $^\dagger$Worcester Polytechnic Institute, Worcester, MA, USA}
\IEEEauthorblockA{\\$^*$\{yuan9, ymo6@vt.edu\}@vt.edu\\$^\dagger$\{pkiaei, pschaumont\}@wpi.edu}
}


%


\maketitle

\begin{abstract}
Lightweight block ciphers have been widely used in applications such as RFID tags, IoTs and network sensors. Among them, with comparable parameters, the Light Encryption Device (LED) block cipher achieves the smallest area. However, implementation of encryption algorithms manifest side-channel leakage, therefore, it is crucial to protect their design against side-channel analysis. In this paper, we present a threshold implementation of the LED cipher which has 64-bit data input and 128-bit key. The presented design splits secret information among multiple shares to achieve a higher security level. We demonstrate that our implementation can protect against first-order power side-channel attack. As a cost, the design area is almost doubled and the maximum operating frequency is degraded by $30\%$. To make our design verifiable, we have also open-sourced our design online.\footnote{https://github.com/Secure-Embedded-Systems/Open-Source-Threshold-Implementation-of-LED-Block-Cipher/} 

\end{abstract}

\begin{IEEEkeywords}
Side-channel Analysis, Side-channel Countermeasure, Threshold Implementation, Open-source.
\end{IEEEkeywords}

%
\IEEEpeerreviewmaketitle

\section{Introduction}
Over the past few years, lightweight cryptography has been largely in demand. Many lightweight block ciphers have been proposed for the highly constrained devices such as RFID tags, sensors in wireless network, small internet-enabled applications with low computing power and implementation area. Improving the encryption efficiency while at the same time protecting the system has become a major challenge in this area and sparked research efforts.  Recently, many encryption ciphers have been developed targeting this problem.
SIMON \cite{shahverdi2015silent}, PRINCE \cite{borghoff2012prince}, TWINE \cite{suzaki2011twine} are only a few of the light-weight block ciphers.

LED \cite{Guo2011} is a lightweight block cipher proposed by Guo et al. at CHES 2011 which, among block ciphers with comparable parameters \cite{mendel2012differential}, achieves the smallest area footprint through its compact hardware implementation. 

Although modern block cipher algorithms such as LED are secure against cryptanalysis, the implementation of an algorithm can still leak information about sensitive parameters, making them susceptible to Side Channel Attacks (SCA). SCAs break cryptosystems by obtaining information leaked from power consumption, electromagnetic radiation, or execution delay. One of the popular SCA schemes is Differential Power Analysis (DPA) \cite{kocher1999differential} which combines information across many power measurements with different known inputs. A widely implemented countermeasure to SCA is masking \cite{blomer2004provably,oswald2005side,ishai2003private,trichina2004small}. However, it has been proven \cite{mangard2005side,moradi2010correlation} that hardware implementations with masking can still be vulnerable to DPA because of glitches and require fresh random values after every nonlinear operation.

To deal with this issue, Nikova et al. proposed a new masking scheme, Threshold Implementation (TI) \cite{nikova2006threshold}. TI is a secret sharing-based countermeasure and is provably secure against first-order SCA attack even in the presence of glitches. 
Moreover, TI is immune to mean power consumption comparison based higher-order attacks.

In this work, we present application of threshold implementation to the lightweight cipher LED. With the advantages of LED and threshold implementation, threshold implementation on LED should be a promising prospect especially in improving the security against SCA in the area of lightweight block cipher application.
Our experimental results show that the proposed implementation can potentially achieve a security level which can efficiently protect against side-channel analysis. On the other hand, with inherent redundancy in threshold implementation, our design almost doubles the area and the maximum operating frequency is degraded by about 30\%.

The rest of the paper is organized as follows. Section~\ref{sec:background} presents the basic concepts of the LED cipher and threshold implementation. Section~\ref{sec:related} reviews the state of the art SCA countermeasures and threshold implementation applications. Section~\ref{sec:ti-led} explains the proposed threshold implementation of LED in detail. The experiment setup and result analysis are given in Section~\ref{sec:results}. The conclusion is discussed in Section~\ref{sec:conclusion}.

\begin{figure}[!t]
\centering
\includegraphics[width=3.2in]{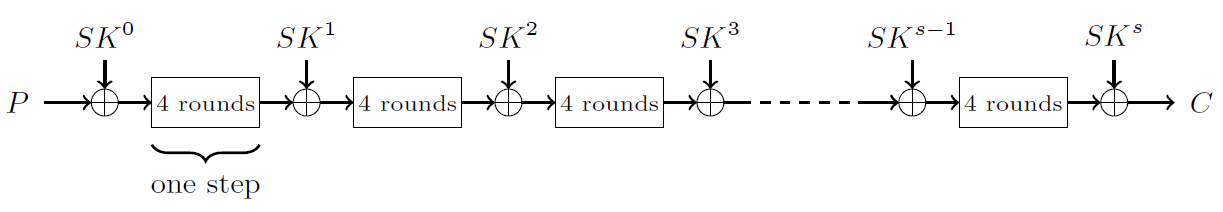}
\caption{Encryption procedure of LED \cite{Guo2011}.}
\label{fig:LED}
\end{figure}

\begin{table}[htbp]
\centering
\caption{Sbox of LED cipher}
\resizebox{9cm}{!} {
{\Large
\label{sbox_table}
\begin{tabular}{|l|l|l|l|l|l|l|l|l|l|l|l|l|l|l|l|l|}
\hline
x	&0	&1	&2	&3	&4	&5	&6	&7	&8	&9	&A	&B	&C	&D	&E	&F    \\ \hline	
S[x] &C	&5	&6	&B	&9	&0	&A	&D	&3	&E	&F	&8	&4	&7	&1	&2   \\ \hline

\end{tabular}
}
}
\end{table}

\section{Background} \label{sec:background}
\subsection{LED}
Light Encryption Device (LED) is a light-weight block cipher that achieves comparable security level to conventional cryptography but at a lower cost. The basic design of LED follows the AES-like design principle. It has many similarities with AES in Sbox, ShiftRows, and MixColumns operations. 

Unlike AES, LED does not have real key scheduling procedure, instead, it initializes round key at the very beginning, and uses that key repeatedly in each round. LED can use three key lengths: 64-bit key, 80-bit key, and 128-bit case. The initial key is generated by specific matrix element setting function. Specifically, for 128-bit key case, the sub-key is divided into two parts and is used alternatively in each round. Therefore, LED has a relatively simple key “schedule” compared to AES. 

As shown in Fig.~\ref{fig:LED}, LED has 8-12 steps depending on its key length and each step has 4 rounds of encryption procedure modifying its cipher states. The procedures are, listed in sequence, the operations: 
AddConstant, SubCells, ShiftRows and MixColumnSerial. Specifically, the number of steps is decided by the encryption key size: for 64-bit key, step number is 8 and for key size 128 bits, step number is 12.
 
LED uses the Sbox from PRESENT \cite{bogdanov2007present}. This sbox has been widely used in many other lightweight cryptographic ciphers. The look up table of Sbox is shown in Table.~\ref{sbox_table}.

\subsection{Serial Implementation of LED}
\begin{figure}[!t]
\centering
\includegraphics[width=3.2in]{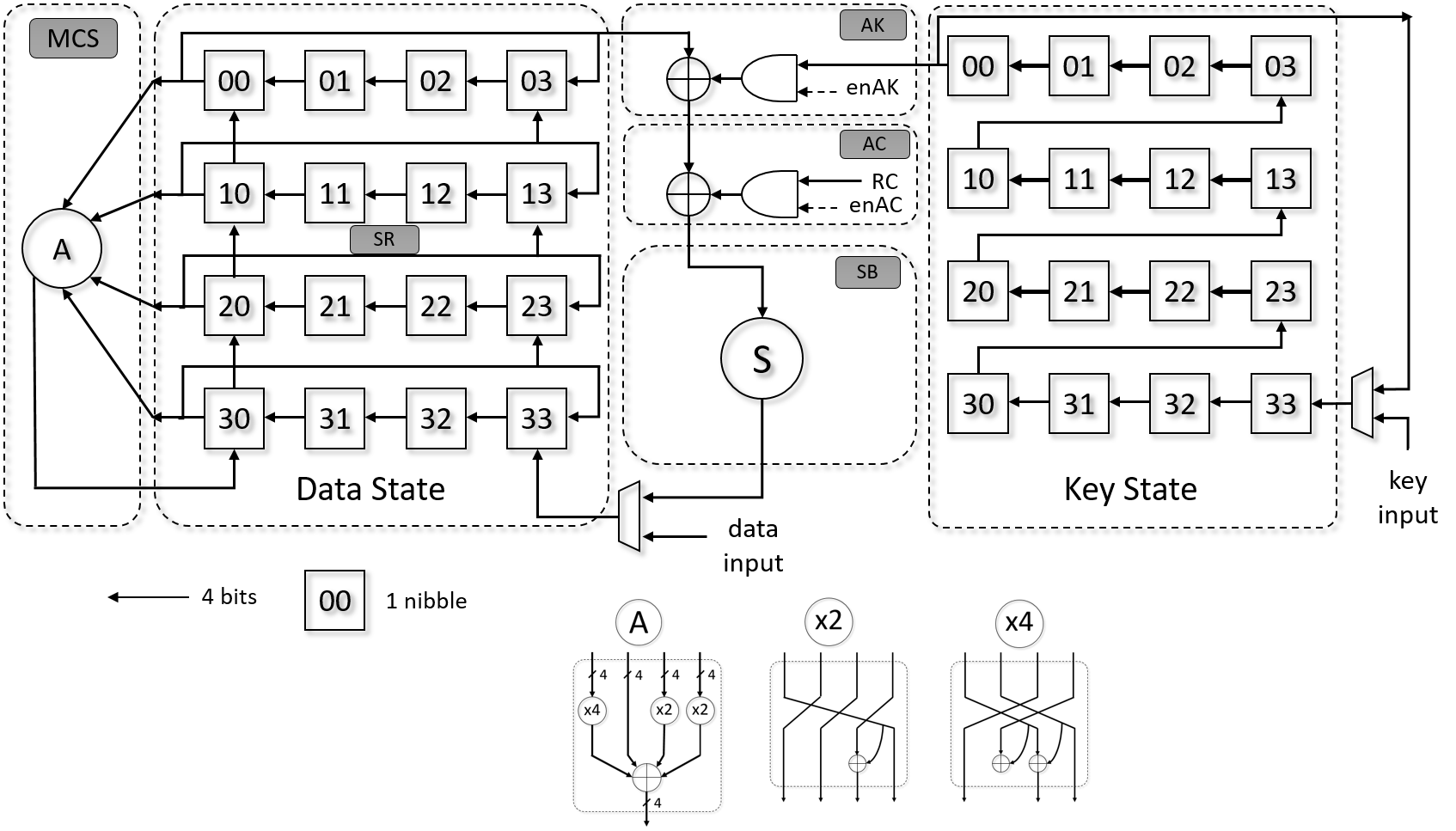}
\caption{Serial Implementation of LED.}
\label{fig:LED_original}
\end{figure}

The nibble-serial implementation of LED which was introduced by Guo et al. \cite{Guo2011} and Farhady et al. \cite{ghalaty2015differential} is shown in Fig. ~\ref{fig:LED_original}. The data state matrix is composed of 16 nibbles. It can compute ShiftRows and MixColumnSerial by serially rotating the matrix row-wise and column-wise. AddRoundKey, AddConstant, and SubBytes are performed by rotating the entire matrix in a serpentine way. Our threshold implementation is developed based on this design. 

\subsection{Threshold Implementation}
Threshold Implementation \cite{nikova2006threshold} is based on secret sharing, threshold cryptography, and multi-party computation protocols. Threshold Implementation splits the secret data and operations into multiple shares. For each share implementation,  three properties should be satisfied:

\subsubsection{Correctness}
The sum of output shares gives the original desired output value.
\subsubsection{Non-completeness}
The function for each output share should be independent of at least one share of each input variable. 
\subsubsection{Uniformity}
Both the input shares and output shares should be uniformly distributed. 

If the implementation of an algorithm satisfies these three properties, all the intermediate values will be independent of the input and output values, thus the correlation is eliminated between output shares and input variables. Moreover, the power consumption or any other characteristics of function will also be independent of input and output. In this way, at any time instant, no information about all shares of secret value is present in the system, even in the presence of glitches. 

\section{Related Work} \label{sec:related}
\subsection{Countermeasures Against SCA}
To provide side-channel attack resistance different masking countermeasures have been introduced at both the algorithm level \cite{blomer2004provably,oswald2005side} and the gate level \cite{ishai2003private,trichina2004small}. Generally, they prevent side-channel leakage by randomizing the sensitive computed intermediates at the algorithm level so that the randomized power consumption does not correlate with the original intermediate values. However, these masking methods do not consider the presence of glitches and need multiple fresh random values along operations.

\subsection{Threshold Implementation Application}
Threshold Implementation has been applied to several block ciphers and Sbox algorithms. $3\times3$ and $4\times4$ Sboxes with 3 and 4 shares are proposed in \cite{bilgin2012threshold}. Moradi et al. \cite{moradi2011pushing} proposed a Threshold Implementation of tower field approach-based AES Sbox that only requires three shares. PRESENT \cite{poschmann2011side}, AES \cite{moradi2011pushing}, Keccak \cite{bertoni2010building,bilgin2013efficient}, Fides \cite{bilgin2013fides}, and SIMON \cite{shahverdi2015silent} with TI protection are also introduced recently. 
Diel et al. described an threshold implementation on LED in 2018 \cite{diehl2018comparing}. However, there is no open-source Threshold Implementation of LED available for researcher to further evaluate-yet. 

\vspace{1.2em}

\section{Implementation Details of LED Threshold Implementation} \label{sec:ti-led}

\begin{figure}[!t]
\centering
\includegraphics[width=2.2in]{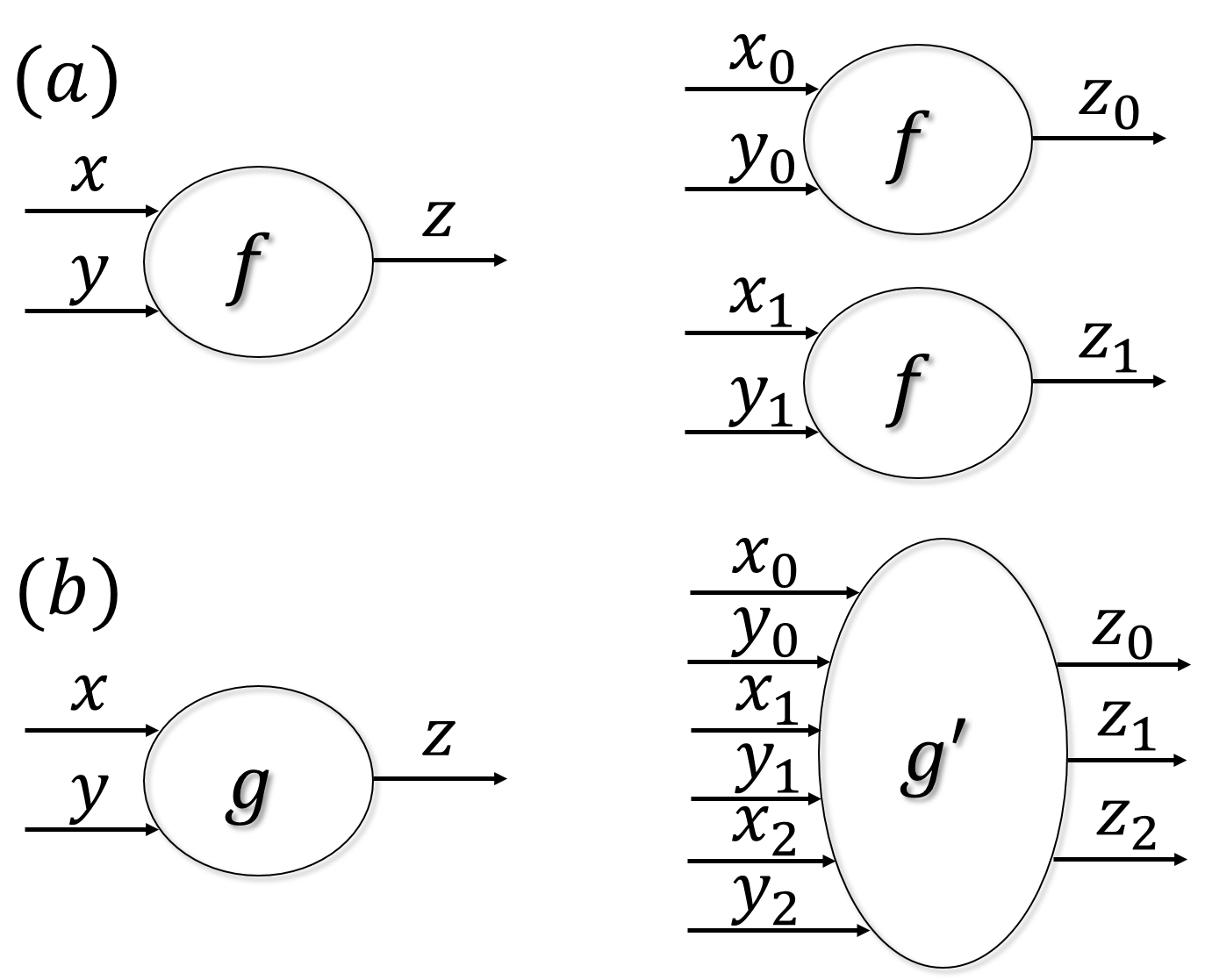}
\caption{Threshold implementation on (a) linear and (b) non-linear operation.}
\label{fig:operation}
\end{figure}

\subsection{Linear and non-linear operations}
The threshold implementation of linear and non-linear operation are different. The threshold implementation of linear operations such as AddRoundKey, AddConstant, ShiftRows and MixColumns are very straightforward. After splitting the input into multiple shares (at least two shares), each share will go through one of the duplicated circuits of the original operation, as shown in Fig.~\ref{fig:operation}. In fact, the requirements of extra copies of hardware contribute to the area and power overhead of threshold implementation. 

\begin{figure}[!t]
\centering
\includegraphics[width=2.5in]{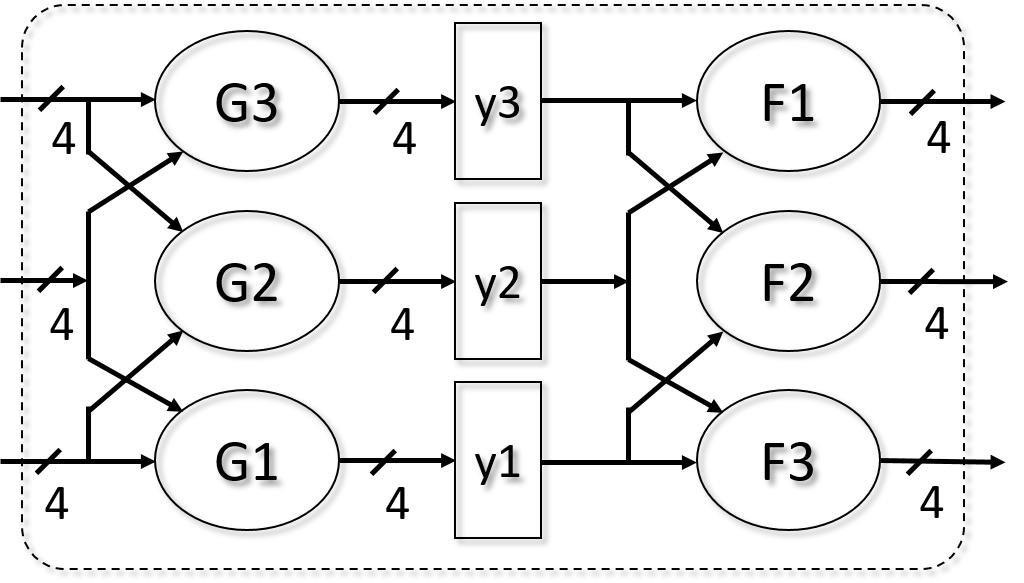}
\caption{Threshold implementation of Sbox used in LED \cite{poschmann2011side}.}
\label{fig:sbox}
\end{figure}

It has been proven \cite{nikova2006threshold} that we need at least three shares to implement a non-linear function $g$. In addition, the new operation $g\prime$ in threshold implementation is not simply duplicating the original operation, as shown in Fig.~\ref{fig:operation}(b). The only non-linear operation in LED is the Sbox operation. Fig.~\ref{fig:sbox} shows the Sbox implementation in LED, it follows the design guideline by Poschmann et al. \cite{poschmann2011side}. The detail of G1, G2, G3, F1, F2 and F3 operations are chosen to satisfy the correctness, non-completeness and uniformity properties required by threshold implementation. The detailed derivation process can be found in their original manuscript \cite{poschmann2011side}.

\begin{figure}[!t]
\centering
\includegraphics[width=3.2in]{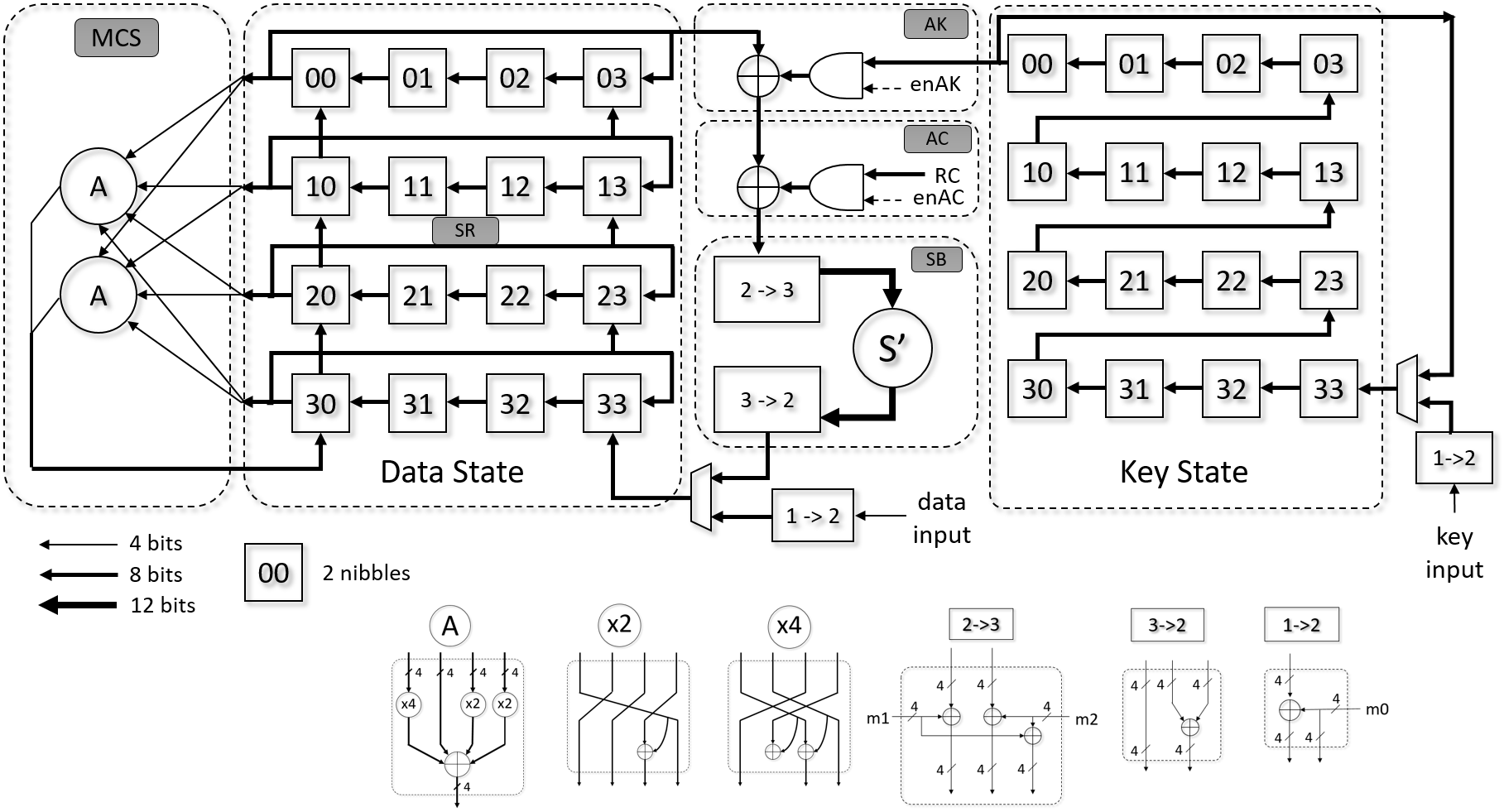}
\caption{Threshold implementation of LED.}
\label{fig:LED_TI}
\end{figure}

\begin{figure}[!t]
\centering
\includegraphics[width=3.2in]{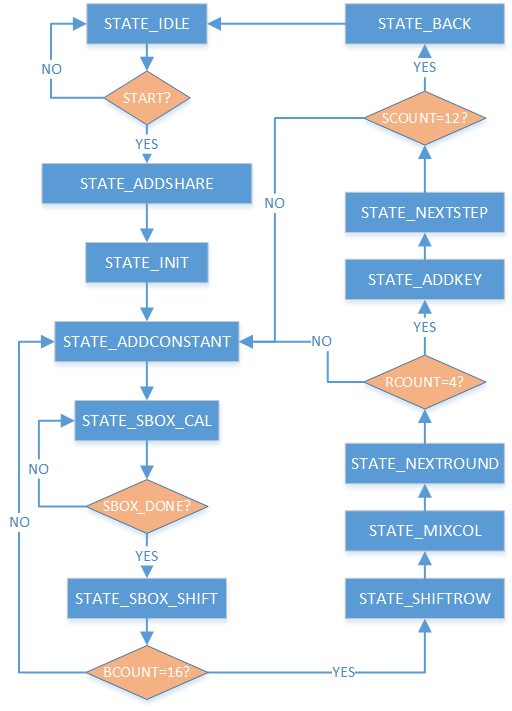}
\caption{Finite state machine of proposed LED threshold implementation.}
\label{fig:fsm}
\end{figure}

\subsection{General data flow and state transition}
The architecture and finite state machine of proposed LED threshold implementation is shown in Fig.~\ref{fig:LED_TI} and Fig.~\ref{fig:fsm}, respectively. The operations of each state are explained as follows.\\

\noindent
$\textbf{STATE\_ADDSHARE:}$ Each data and key nibble is split into two nibble shares while loading into the data state matrix in a byte-serial manner. Therefore, each position of data state matrix stores a data byte. As shown in Fig. ~\ref{fig:LED_TI}, the $1\rightarrow2$ add share operation is achieved using a 4 bit random number m0. The AddShare operation takes 16 cycles to load in all the data and key nibbles.\\

\noindent
\textbf{$\textbf{STATE\_INIT:}$} In this state, the initial AddRoundKey operation (AK in Fig.~\ref{fig:LED_TI}) ($SK^0$) is performed. The enAK is set to high to enable the round key addition. The two data shares in position 00 are XORed with the two key shares in position 00.\\

\noindent
$\textbf{STATE\_ADDCONSTANT:}$ As shown in the AC part of Fig.~\ref{fig:LED_TI}, the data shares are XORed with the round constant RC with enAC being set to high. Note that the 4 bit round constant also needs to be split into two shares. Then, state jumps to $STATE\_SBOX\_CAL$.\\

\noindent
$\textbf{STATE\_SBOX\_CAL:}$ The SubBytes operation (SB in Fig.~\ref{fig:LED_TI}) is performed in this state. As mentioned before, due to the non-linearity of the Sbox, three data shares are needed. As shown in Fig.~\ref{fig:LED_TI}, the data shares are increased to three using the $2\rightarrow3$ operation before entering the Sbox and reduced to two using the $3\rightarrow2$ operation when exiting the Sbox for later linear operations. This state takes three cycles. Then, state jumps to $STATE\_SBOX\_SHIFT$.\\

\noindent
$\textbf{STATE\_SBOX\_SHIFT:}$ In this state, the two data shares are fed back into the data state matrix in position 33 in a byte serial manner. Each data byte in the matrix shifts forward in a serpentine way. Then, the state jumps to $STATE\_ADDCONSTANT$ until all 16 data bytes finish the AddConstant and SubBytes operation (BCOUNT=16). Therefore, in total $4\times16=64$ cycles are required to finish. Then, the state jumps to $STATE\_SHIFTROW$.\\

\noindent
$\textbf{STATE\_SHIFTROW:}$ The ShiftRow operation (SR in Fig.~\ref{fig:LED_TI}) will be performed in this state as well as the update of round constant. Then, the state jumps to $STATE\_MIXCOL$.\\

\noindent
$\textbf{STATE\_MIXCOL:}$ The MixColumn operation (MCS in Fig.~\ref{fig:LED_TI}) is applied on the column 0 and rotates row-wise after it finishes. For the leftmost column, the MixColumn operation is performed on the byte 0 and then rotates column-wise. Therefore, in total $4\times4=16$ cycles are needed to finish. Since MixColumn is a linear operation, we can simply duplicate the operation. Then, the state jumps to $STATE\_NEXTROUND$.\\

\noindent
$\textbf{STATE\_NEXTROUND:}$ In this state, if we have already completed all four rounds (RCOUNT=4), state jumps to the start of next step which is $STATE\_ADDKEY$, otherwise, state jumps to $STATE\_ADDCONSTANT$.\\

\noindent
$\textbf{STATE\_ADDKEY:}$ The AddRoundKey will be performed in this state. If this is the end of last step (SCOUNT=12), the encryption finishes and we jump to $STATE\_BACK$. Otherwise, we continue to the start of next step, which is $STATE\_ADDCONSTANT$.\\

\noindent
$\textbf{STATE\_BACK:}$ In this state, for each data byte or two data nibble shares, the $2\rightarrow1$ operation will be performed to get the encrypted data nibble. The operation is done in byte serial manner and takes 16 cycles to finish. Afterwards, it will jump to $STATE\_IDLE$. \\

\begin{figure}[!t]
\centering
\includegraphics[width=3.2in]{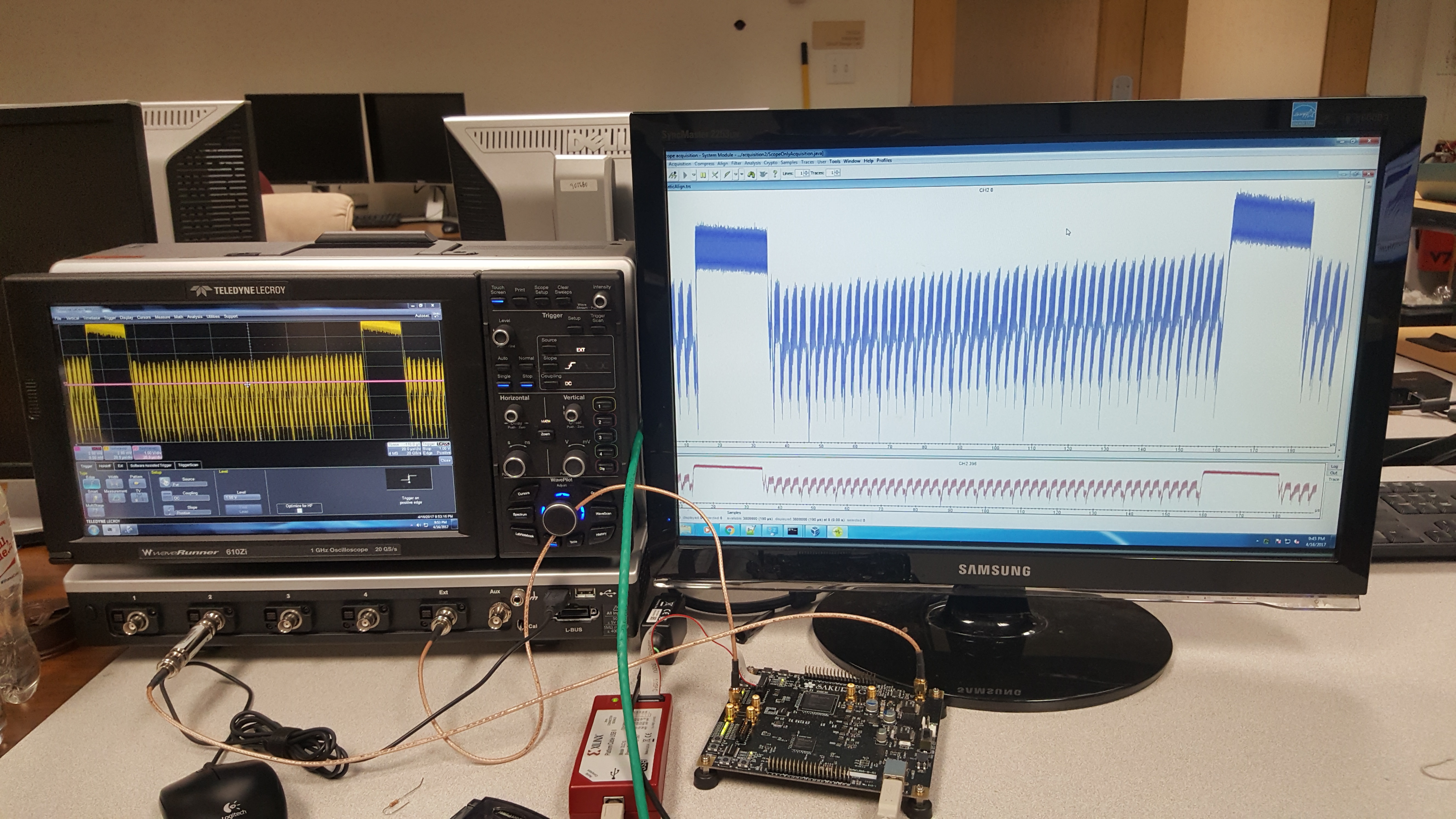}
\caption{SAKURA-G board with Spartan™-6 FPGA and TELEDYNE LECROY WaveRunner 6 Zi HRO Oscilloscope.}
\label{fig:setup}
\end{figure}

\begin{figure}[!t]
\centering
\includegraphics[width=3.2in]{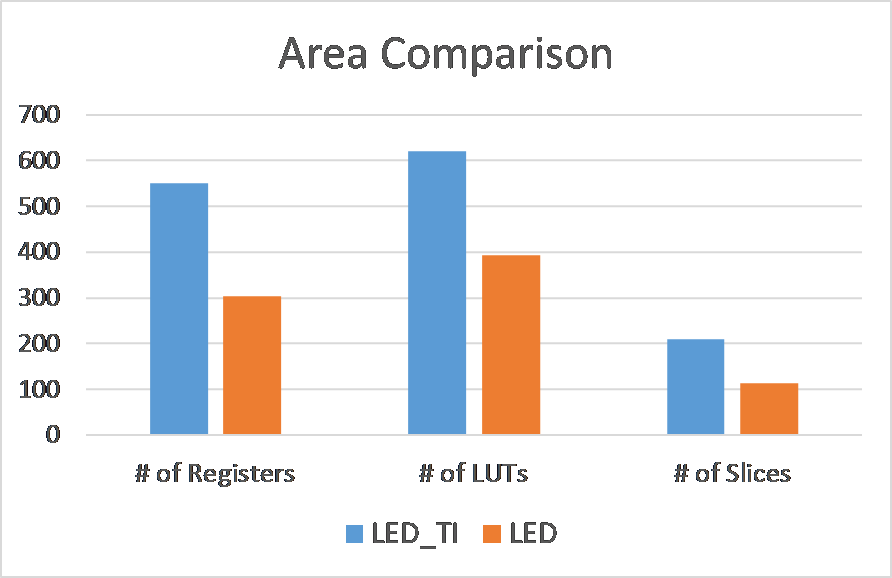}
\caption{Area comparison of $LED\_TI$ (threshold implementation) and $LED$ (unprotected).}
\label{fig:area}
\end{figure}

\begin{figure}[!t]
\centering
\includegraphics[width=3.2in]{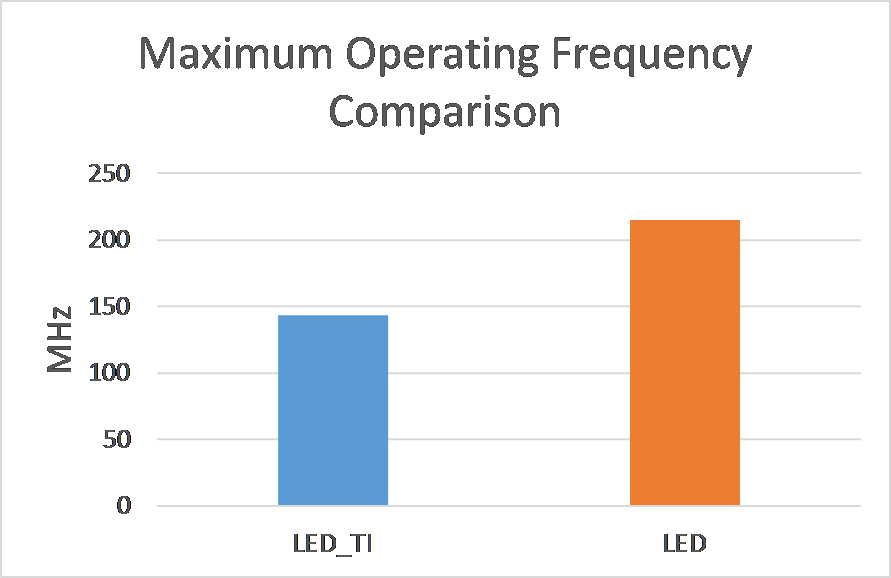}
\caption{Maximum operating frequency comparison of $LED\_TI$ (threshold implementation) and $LED$ (unprotected).}
\label{fig:freq}
\end{figure}



\input{Experiment_Result}

\input{conclusion}
\input{ref}

\vspace{1.2em}



%



\end{document}

%% file: Experiment_Result.tex
\section{Experiment and Expected Results} \label{sec:results}
\subsection{Experimental setup}


In order to evaluate our Threshold Implementation of LED, we add it to Skiva\footnote{https://github.com/Secure-Embedded-Systems/Skiva}\cite{kiaei2020custom} as a coprocessor. Skiva is an extension of the 32bit SPARC V8 ISA. The connection between the Skiva processor and the LED coprocessor is through an Advanced Peripheral Bus (APB). We communicate with the LED accelerator by sending data to its memory-mapped registers. We assign memory addresses for control register, key registers, and plaintext registers within the LED coprocessor. The SoC implements a 32bit bus with a word-addressed memory. Therefore, in our LED-128 implementation, we assign two memory addresses for the plaintext registers (64 bits) and four memory addresses for the key registers (128 bits).
The control register contains two bits; One for resetting the LED core and one for starting the encryption.
Furthermore, to make our experimental evaluation as precise as possible, we add a trigger signal to the LED core indicating when the encryption is taking place. Triggering from the LED core helps in precise alignment of our power measurements. 

We implement our design on Xilinx Spartan-6 FPGA on the SAKURA-G board. 
While running LED encryption on the main FPGA, we collect power traces for the encryption procedure and use it for power analysis. The basic setup is shown in Fig.~\ref{fig:setup}. 
TELEDYNE LECROY WaveRunner 6 Zi HRO Oscilloscope is used for power trace collection. 
Power trace sets collected from the oscilloscope will be sent to PC for further analysis.

\begin{figure}[h]
\centering
\includegraphics[width=\columnwidth]{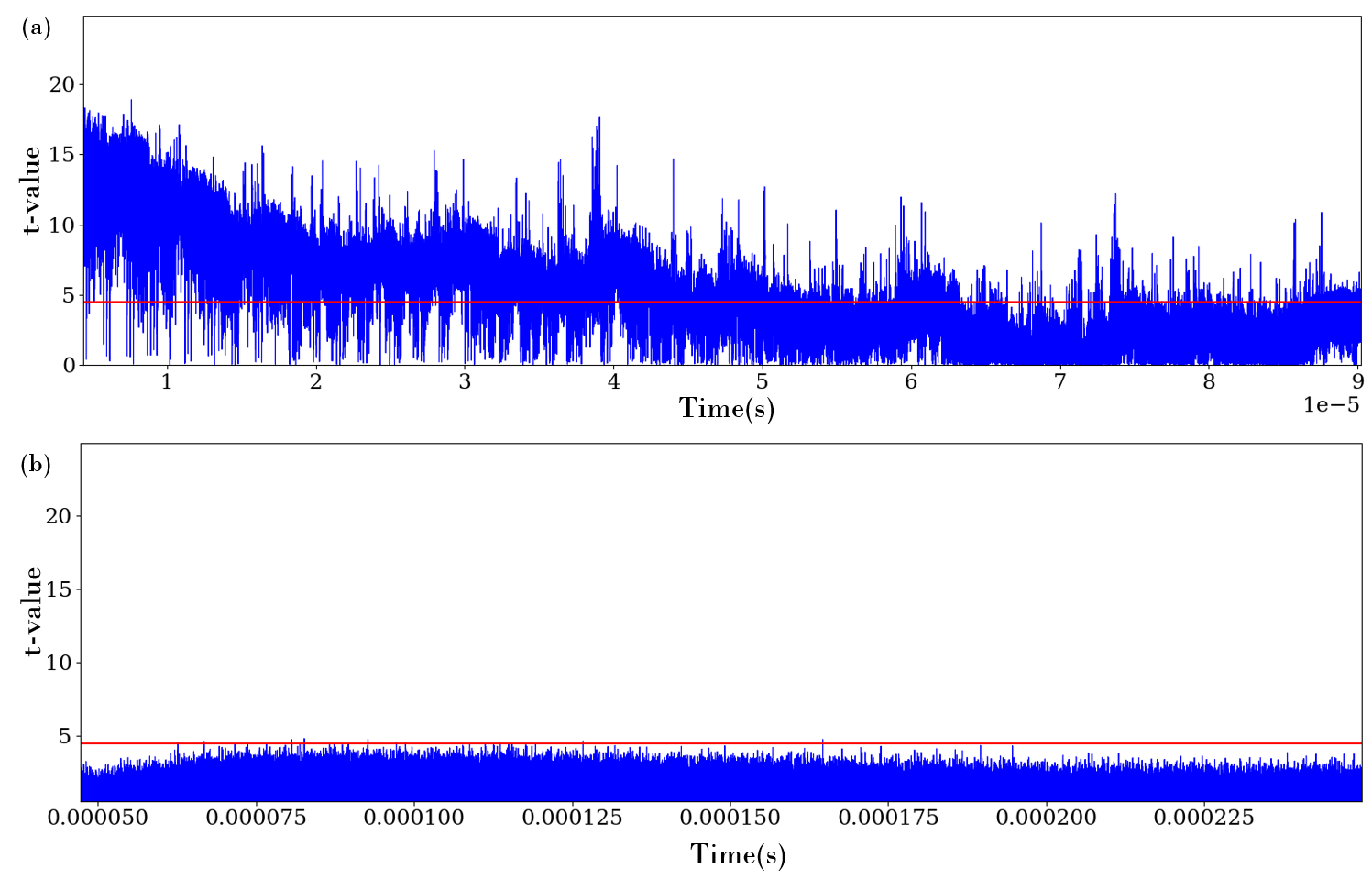}
\caption{T-value comparison of $LED\_TI$ and $LED$.}
\label{fig:t-comp}
\end{figure}

The area and maximum operating frequency are obtained after synthesizing our design using Spartan-6 FPGA, as shown in Fig.~\ref{fig:area} and Fig.~\ref{fig:freq}. $LED\_TI$ denotes our proposed LED threshold implementation, while $LED$ is the original unprotected design. It can be seen that $LED\_TI$ almost doubles the area because of the extra copies of hardware for other shares. The maximum frequency is reduced by about $30\%$. This is due to the more complicated computations operated on multiple shares.

To further evaluate the security of the proposed Threshold implementation design on increasing side-channel resistance, we applied TVLA \cite{gilbert2011testing} on 50k collected traces for both $LED\_TI$ and $LED$; As shown in Fig.~\ref{fig:t-comp}, a dramatic decrease of t-value can be observed for $LED\_TI$  compared to unprotected LED. This indicates that our Threshold Implementation design can significantly reduce the side-channel leakage of LED.

%% file: conclusion.tex
\section{Conclusion} \label{sec:conclusion}

In this paper, we designed a byte-serial threshold implementation of LED block cipher to mitigate first order power side-channel analysis attacks. In our understanding, this is the first open-source threshold implementation of LED cipher. 
In the aspect of cost, our design almost doubles the area and reduces the speed by $30\%$ in comparison with a reference unprotected nibble-serial LED design. We collected power consumption traces for both the reference (unprotected) and threshold implementation (protected) and we demonstrate that our proposed implementation of Threshold Implementation can effectively protect LED from side-channel leakage. 

%% file: ref.tex
\bibliographystyle{ieeepes}
\newlength{\bibitemsep}\setlength{\bibitemsep}{0\baselineskip plus .05\baselineskip minus .05\baselineskip}
\newlength{\bibparskip}\setlength{\bibparskip}{0pt}
\let\oldthebibliography\thebibliography
\renewcommand\thebibliography[1]{%
  \oldthebibliography{#1}%
  \setlength{\parskip}{\bibitemsep}%
  \setlength{\itemsep}{\bibparskip}%
}
\bibliography{semsref}